\documentclass{ws-procs9x6-cpt19}
\usepackage{makecell}
\begin{document}
\newcommand{\refeq}[1]{(\ref{#1})}
\def\etal {{\it et al.}}

\title{Constraining Dimension-Six Nonminimal Lorentz-Violating Electron--Nucleon
	Interactions with EDM Physics}

\author{Jonas B.\ Araujo,$^1$ A.H.\ Blin,$^2$ Marcos Sampaio,$^{3}$, and Manoel M.\ Ferreira Jr$^1$}

\address{$^1$Departamento de F\'{\i}sica, Universidade Federal do Maranh\~{a}o,\\
Campus Universit\'{a}rio do Bacanga, S\~{a}o Lu\'{\i}s - MA, 65080-805, Brazil}
\address{$^2$CFisUC, Department of Physics, University of Coimbra,\\
3004-516 Coimbra, Portugal}
\address{$^3$CCNH, Universidade Federal do ABC, Santo Andr\'e - SP, 09210-580, Brazil}

\begin{abstract}
Electric dipole moments of atoms 
can arise from P-odd and T-odd electron--nucleon couplings. 
This work studies a general class of dimension-six electron--nucleon interactions
mediated by Lorentz-violating tensors 
of ranks ranging from $1$ to $4$. 
The possible couplings are listed 
as well as their behavior under C, P, and T, 
allowing us to select the couplings compatible with electric-dipole-moment physics. 
The unsuppressed contributions of these couplings 
to the atom's hamiltonian 
can be read as equivalent to an electric dipole moment. 
The Lorentz-violating coefficients' magnitudes 
are limited using electric-dipole-moment measurements at the levels of 
$3.2\times10^{-31}\text{(eV)}^{-2}$ or $1.6\times10^{-33}\text{(eV)}^{-2}$.
\end{abstract}

\bodymatter

\section{Introduction}
Electric dipole moments (EDMs) are excellent probes for 
violations of discrete symmetries 
and for physics beyond the Standard Model.\cite{EDM3}
EDM terms violate both parity (P) and time-reversal (T) symmetry, 
while preserving charge conjugation (C),
assuming the CPT theorem holds. 
An atom's EDM could arise from intrinsic properties of the electrons and/or nucleus, or from P- and T-odd electron--nucleon ($e$--$N$) couplings. 
EDM phenomenology can also arise in a Lorentz-violation (LV) scenario 
addressed within the framework of the Standard-Model Extension (SME).\cite{Colladay} 
LV generates CP violation and EDMs 
via radiative corrections\cite{Haghig} 
or at tree level via dimension-five nonminimal couplings,\cite{Pospelov2008,UFMA} 
and may also yield nuclear-EDM corrections 
to the Schiff moment.\cite{Jonas3} 
Nonminimal couplings have been of great interest in recent years.\cite{NMC}
In this work, 
we investigate a class of dimension-six LV $e$--$N$ couplings, 
composed of rank-$1$ to rank-$4$ background tensors, 
first proposed in Ref.\ \refcite{LVArbDim}, 
and the possible generation of atomic EDMs.\cite{Yamanaka}

\section{Nonminimal $e$--$N$ Lorentz-violating couplings}
\label{section2}

The simplest couplings involve a rank-$1$ LV tensor $(k_{XX})_{\mu}$ 
with an effective lagrangian of the form 
$
\mathcal{L}_{\text{LV}}=(k_{XX})_{\mu}\left[  \left(  \bar{N}\,\Gamma_{1}\,N\right)
\left(  \bar{\psi}\,\Gamma_{2}\,\psi\right)  \right]  ^{\mu},
$ 
where the upper $\mu$ 
belongs either to $\Gamma_{1}$ or $\Gamma_{2}$. 
The subscript $XX$ 
labels the type of fermion bilinear: 
scalar ($S$), 
pseudoscalar ($P$), 
vector ($V$), 
axial vector ($A$), 
and tensor ($T$), 
accounting for the $16$ linearly independent $4\times4$ matrices. 
The operators $\Gamma_{1,2}$ must be combinations of Dirac matrices,
given here as
\begin{equation}
\alpha^{i}=\left(
\begin{array}
[c]{cc}%
0 & \sigma^{i}\\
\sigma^{i} & 0
\end{array}
\right)  ,\ \ \Sigma^{k}=\left(
\begin{array}
[c]{cc}%
\sigma^{k} & 0\\
0 & \sigma^{k}%
\end{array}
\right)  ,\ \ \gamma^{0}=\left(
\begin{array}
[c]{cc}%
1 & 0\\
0 & -1
\end{array}
\right)  ,\ \ \gamma^{5}=\left(
\begin{array}
[c]{cc}%
0 & 1\\
1 & 0
\end{array}
\right)  ,\label{Defsgamma}%
\end{equation}
and $\gamma^{i}=\gamma^{0}\alpha^{i}$, 
$\sigma^{\mu\nu}=i\left[  \gamma^{\mu},\gamma^{\nu}\right]  /2$, 
with $\sigma^{0j}=i\alpha^{j}$, 
$\sigma^{ij}=\epsilon_{ijk}\Sigma^{k}$. 
As we are
interested in EDM behavior, 
we select the P-odd and T-odd components. 
The rank-$1$ couplings are listed in Table \ref{TableI}, 
in which ``S" and ``NS" mean
``suppressed" and ``not suppressed," 
respectively, 
in the nucleon's nonrelativistic limit.
Couplings of higher ranks are shown in Tables \ref{TableII} to \ref{TableIV}.

\begin{table}[h]
\tbl{General CPT-odd couplings with a rank-$1$ LV tensor.}
{\begin{tabular}{@{}cccc@{}}\toprule
Coupling &  P-odd, T-odd piece  & NRL & EDM \\
\colrule
$\left(k_{SV}\right) _{\mu}(\bar{N}N)(\bar{\psi}\gamma^{\mu}\psi)$ &
$\left(k_{SV}\right) _{i}(\bar{N}N)(\bar{\psi}\gamma^{i}\psi)$ & NS & yes \\

$\left(k_{VS}\right) _{\mu}(\bar{N}\gamma^{\mu}N)(\bar{\psi}\psi)$  &
$\left(k_{VS}\right) _{i}(\bar{N}\gamma^{i}N)(\bar{\psi}\psi)$ & S & --\\

$\left(k_{VP}\right)  _{\mu}(\bar{N}\gamma^{\mu}N)(\bar{\psi}i\gamma_{5}\psi)$ &
 $\left(k_{VP}\right) _{0}(\bar{N}\gamma^{0}N)(\bar{\psi}i\gamma_{5}\psi)$ & NS & yes \\	

$\left(k_{PV}\right)  _{\mu}(\bar{N}i\gamma_{5}N)(\bar{\psi}\gamma^{\mu}\psi)$ &
$\left(k_{PV}\right) _{0}(\bar{N}\gamma_{5}N)(\bar{\psi}\gamma^{0}\psi)$ & S & --\\
$\left(k_{SA}\right)  _{\mu}(\bar{N}N)(\bar{\psi}\gamma^{\mu}\gamma_{5}\psi)$  &
none & -- & --\\
$\left(k_{AS}\right)  _{\mu}(\bar{N}\gamma^{\mu}\gamma_{5}N)(\bar{\psi}\psi)$ &none & -- & --\\
$\left(k_{PA}\right)  _{\mu}(\bar{N}i\gamma_{5}N)(\bar{\psi}\gamma^{\mu}
\gamma_{5}\psi)$   & none & -- & -- \\
$\left(k_{AP}\right)  _{\mu}(\bar{N}\gamma^{\mu}\gamma_{5}N)(\bar{\psi}
i\gamma_{5}\psi)$   & none & -- & --\\
\botrule
\end{tabular}
}
\label{TableI}
\end{table}

The EDM contribution, 
in the nonrelativistic limit for the nucleons, 
is calculated via atomic parity nonconservation methods\cite{LeptonEDM} as
\begin{equation}
\frac{\Delta E}{E_{z}}=2e\Re\left[  \langle\psi_{0}|H_{P,T}|\eta
\rangle\right]  \equiv d_{\text{equiv}},\label{EDMequiv}%
\end{equation}
where $H_{P,T}$ is the coupling's hamiltonian contribution for the electron, 
and $\vert \eta\rangle,\vert \psi_0\rangle$---which have opposite parity---correspond to
\begin{equation}
\left(\psi_0\right)^{l}=\left(
\begin{array}
[c]{c}%
\frac{i}{r}G_{l,J=\frac{1}{2}}(r)\phi_{\frac{1}{2},\frac{1}{2}}^{l}\\
\frac{1}{r}F_{l,J=\frac{1}{2}}(r)  \boldsymbol{\sigma}\cdot
\hat{\boldsymbol{r}}  \phi_{\frac{1}{2},\frac{1}{2}}^{l}%
\end{array}
\right)  ,\ 
\eta^{l}=\left(
\begin{array}
[c]{c}%
\frac{i}{r}G_{l,J=\frac{1}{2}}^{S}(r)\phi_{\frac{1}{2},\frac{1}{2}}^{l}\\
\frac{1}{r}  F_{l,J=\frac{1}{2}}^{S}(r)\boldsymbol{\sigma}\cdot
\hat{\boldsymbol{r}}  \phi_{\frac{1}{2},\frac{1}{2}}^{l}%
\end{array}
\right),
\end{equation}
where
$
\phi_{\frac{1}{2},\frac{1}{2}}^{l=0}
=(Y_{0}^{0},0)^T$
and 
$ \phi_{\frac{1}{2},\frac{1}{2}}^{l=1}
=(\sqrt{\frac{1}{3}}Y_{1}^{0},
-\sqrt{\frac{2}{3}}Y_{1}^{1})^T$.
While $\left(\psi_0\right)^{l}$ obeys Dirac's equation for a central potential, $\eta^{l}$ obeys Sternheimer's equation.\cite{LeptonEDM}

\begin{table}
		\tbl{General CPT-even couplings with a rank-$2$ LV tensor.}
		{\begin{tabular}{@{}cccc@{}}\toprule
		Coupling &  P-odd, T-odd piece  & NRL & EDM \\\colrule
			$\left(k_{VV}\right)_{\mu\nu}\left(\bar{N}\gamma^{\mu}N\right)\left(\bar{\psi}\gamma^{\nu}\psi\right)$ &   \makecell{$\left(k_{VV}\right)_{i0}\left(\bar{N}\gamma^{i}N\right)\left(\bar{\psi}\gamma^{0}\psi\right)$ \\
			$\left(k_{VV}\right)_{0i}\left(\bar{N}\gamma^{0}N\right)\left(\bar{\psi}\gamma^{i}\psi\right)$} & \makecell{S \\ NS} & \makecell{-- \\ yes}\tabularnewline
		
			$\left(k_{AV}\right)_{\mu\nu}\left(\bar{N}\gamma^{\mu}\gamma_{5}N\right)\left(\bar{\psi}\gamma^{\nu}\psi\right)$   & none & -- & --\tabularnewline
		
			$\left(k_{VA}\right)_{\mu\nu}\left(\bar{N}\gamma^{\mu}N\right)\left(\bar{\psi}\gamma^{\nu}\gamma_{5}\psi\right)$   & none & -- & --\tabularnewline
		
			$\left(k_{AA}\right)_{\mu\nu}\left(\bar{N}\gamma^{\mu}\gamma_{5}N\right)\left(\bar{\psi}\gamma^{\nu}\gamma_{5}\psi\right)$   & \makecell{$\left(k_{AA}\right)_{0i}\left(\bar{N}\gamma^{0}\gamma_{5}N\right)\left(\bar{\psi}\gamma^{i}\gamma_{5}\psi\right)$\\
			$\left(k_{AA}\right)_{i0}\left(\bar{N}\gamma^{i}\gamma_{5}N\right)\left(\bar{\psi}\gamma^{0}\gamma_{5}\psi\right)$} & \makecell{S \\ NS} & \makecell{-- \\ yes} \tabularnewline
			
			$\left(k_{TS}\right)_{\mu\nu}\left(\bar{N}\sigma^{\mu\nu}N\right)\left(\bar{\psi}\psi\right)$   & none & -- & --\tabularnewline

			$\left(k_{TP}\right)_{\mu\nu}\left(\bar{N}\sigma^{\mu\nu}N\right)\left(\bar{\psi}i\gamma_{5}\psi\right)$   & none & -- & --\tabularnewline

			$\left(k_{ST}\right)_{\mu\nu}\left(\bar{N}N\right)\left(\bar{\psi}\sigma^{\mu\nu}\psi\right)$ &   none & -- & --\tabularnewline
			
			$\left(k_{PT}\right)_{\mu\nu}\left(\bar{N}i\gamma_{5}N\right)\left(\bar{\psi}\sigma^{\mu\nu}\psi\right)$   & none & -- & --\tabularnewline
\botrule
\end{tabular}
}
\label{TableII}
\end{table}
	
\begin{table}
	\tbl{General CPT-odd couplings with a rank-$3$ LV tensor.}
	{\begin{tabular}{@{}cccc@{}}\toprule
			Coupling &  P-odd, T-odd piece  & NRL & EDM \\\colrule
		$\left(k_{VT}\right)_{\alpha\mu\nu}(\bar{N}\gamma^{\alpha}N)(\bar{\psi}\sigma^{\mu\nu}\psi)$ &   none & -- & --\tabularnewline
	
		$\left(k_{AT}\right)_{\alpha\mu\nu}(\bar{N}\gamma^{\alpha}\gamma_{5}N)(\bar{\psi}\sigma^{\mu\nu}\psi)$ &   \makecell{$\left(k_{AT}\right)_{0ij}(\bar{N}\gamma^{0}\gamma_{5}N)(\bar{\psi}\sigma^{ij}\psi)$
			\\
			$\left(k_{AT}\right)_{i0j}(\bar{N}\gamma^{i}\gamma_{5}N)(\bar{\psi}\sigma^{0j}\psi)$
			\\
			$\left(k_{AT}\right)_{ij0}(\bar{N}\gamma^{i}\gamma_{5}N)(\bar{\psi}\sigma^{j0}\psi)$} & \makecell{S \\ NS \\ NS} & \makecell{-- \\ yes \\ yes}\tabularnewline

		$\left(k_{TV}\right)_{\alpha\mu\nu}(\bar{N}\sigma^{\mu\nu}N)(\bar{\psi}\gamma^{\alpha}\psi)$ &   none & -- & --\tabularnewline

		$\left(k_{TA}\right)_{\alpha\mu\nu}(\bar{N}\sigma^{\mu\nu}N)(\bar{\psi}\gamma^{\alpha}\gamma_{5}\psi)$ &   \makecell{$\left(k_{TA}\right)_{0ij}(\bar{N}\sigma^{ij}N)(\bar{\psi}\gamma^{0}\gamma_{5}\psi)$
			\\
			$\left(k_{TA}\right)_{i0j}(\bar{N}\sigma^{0j}N)(\bar{\psi}\gamma^{i}\gamma_{5}\psi)$
			\\
			$\left(k_{TA}\right)_{ij0}(\bar{N}\sigma^{j0}N)(\bar{\psi}\gamma^{i}\gamma_{5}\psi)$} & \makecell{NS \\ S \\ S} & \makecell{yes \\ -- \\ --}\tabularnewline
		
	\botrule
\end{tabular}
}
\label{TableIII}
\end{table}

\begin{table}
	\tbl{General CPT-even couplings with a rank-$4$ LV tensor.}
	{\begin{tabular}{@{}cccc@{}}\toprule
			Coupling &  P-odd, T-odd piece  & NRL & EDM \\\colrule
		$\left(k_{TT}\right)_{\alpha\beta\mu\nu}(\bar{N}\sigma^{\alpha\beta}N)(\bar{\psi}\sigma^{\mu\nu}\psi)$   & \makecell{$\left(k_{TT}\right)_{0ijk}(\bar{N}\sigma^{0i}N)(\bar{\psi}\sigma^{jk}\psi)$
			\\
			$\left(k_{TT}\right)_{i0jk}(\bar{N}\sigma^{i0}N)(\bar{\psi}\sigma^{jk}\psi)$
			\\
			$\left(k_{TT}\right)_{ij0k}(\bar{N}\sigma^{ij}N)(\bar{\psi}\sigma^{0k}\psi)$
			\\
			$\left(k_{TT}\right)_{ijk0}(\bar{N}\sigma^{ij}N)(\bar{\psi}\sigma^{k0}\psi)$} & \makecell{S \\ S \\ NS \\ NS} & \makecell{-- \\ -- \\ yes \\ yes} \tabularnewline
		
	\botrule
\end{tabular}
}
\label{TableIV}
\end{table}

By evaluating $d_{\text{equiv}}$ for each coupling via Eq.\ \refeq{EDMequiv} 
and performing a sidereal analysis,\cite{UFMA} 
one can set upper bounds on the LV coefficients 
using numerical estimates on the thallium atom 
and the experimental limit on the electron's EDM.\cite{EDMnature2018} 
A list of time-averaged upper bounds is given in Table \ref{TableBounds}. 
More information can be found in Ref.\ \refcite{added}.

\begin{table}
	\tbl{List of bounds on the LV tensors of ranks ranging from $1$ to $4$.}
	{\begin{tabular}{@{}lc@{}}\toprule
		Component & Upper bound \\\colrule
			$|\left(k_{VP}\right)_{0}^{\text{(Sun)}}|$ & $1.6\times10^{-33}(\text{eV})^{-2}$\tabularnewline
		
			$|\frac{1}{4}\left[\left(k_{AT}\right)^{\text{(Sun)}}_{101}+\left(k_{AT}\right)^{\text{(Sun)}}_{202}-2\left(k_{AT}\right)^{\text{(Sun)}}_{303}\right]\sin2\chi\ |$ & $3.2\times10^{-31}(\text{eV})^{-2}$\tabularnewline
		
			$|\left[-\left(k_{AT}\right)^{\text{(Sun)}}_{102}+\left(k_{AT}\right)^{\text{(Sun)}}_{201}\right]\sin\chi|$ & $3.2\times10^{-31}(\text{eV})^{-2}$\tabularnewline
	
			$|\left[\frac{1}{2}\left( \left(k_{AT}\right)^{\text{(Sun)}}_{101}+\left(k_{AT}\right)^{\text{(Sun)}}_{202} \right)\sin^2\chi+ \left(k_{AT}\right)^{\text{(Sun)}}_{303}\cos^2\chi \right]|$ & $3.2\times10^{-31}(\text{eV})^{-2}$\tabularnewline

			$|\frac{1}{4}\left[\left(K_{TT}\right)^{\text{(Sun)}}_{011}+\left(K_{TT}\right)^{\text{(Sun)}}_{022}-2\left(K_{TT}\right)^{\text{(Sun)}}_{033}\right]\sin 2 \chi |$ & $3.2\times10^{-31}(\text{eV})^{-2}$\tabularnewline
		
			$|\left[\left(K_{TT}\right)^{\text{(Sun)}}_{012}-\left(K_{TT}\right)^{\text{(Sun)}}_{021}\right]\sin\chi |$ & $3.2\times10^{-31}(\text{eV})^{-2}$\tabularnewline
		
			$|\left[\frac{1}{2}\left( \left(K_{TT}\right)^{\text{(Sun)}}_{011}+\left(K_{TT}\right)^{\text{(Sun)}}_{022} \right)\sin^2\chi+ \left(K_{TT}\right)^{\text{(Sun)}}_{033}\cos^2\chi \right] |$ & $3.2\times10^{-31}(\text{eV})^{-2}$\tabularnewline
		\botrule
	\end{tabular}
}
	\label{TableBounds}
\end{table}
\section*{Acknowledgments}
We are grateful to CAPES, CNPq, FCT Portugal, and FAPEMA.


\begin{thebibliography}{xx}


\bibitem {EDM3}
J.\ Engel, M.J.\ Ramsey-Musolf, and U.\ van Kolck,
Prog.\ Part.\ Nucl.\ Phys.\  {\bf 71}, 21 (2013);
T.\ Chupp and M.J.\ Ramsey-Musolf,
Phys.\ Rev.\ C {\bf 91}, 035502 (2015);
N.~Yamanaka \etal,
Eur.\ Phys.\ J.\ A {\bf 53}, 54 (2017).

\bibitem {Colladay}
D.\ Colladay and V.A.\ Kosteleck\'y,
Phys.\ Rev.\ D {\bf 55}, 6760 (1997);
Phys.\ Rev.\ D {\bf 58}, 116002 (1998);
S.R.\ Coleman and S.L.\ Glashow,
Phys.\ Rev.\ D {\bf 59}, 116008 (1999).

\bibitem {Haghig}
M.\ Haghighat, I.\ Motie, and Z.\ Rezaei,
Int.\ J.\ Mod.\ Phys.\ A {\bf 28}, 1350115 (2013).

\bibitem {Pospelov2008}
P.A.\ Bolokhov, M.\ Pospelov, and M.\ Romalis,
Phys.\ Rev.\ D {\bf 78}, 057702 (2008).

\bibitem {UFMA}
J.B.\ Araujo, R.\ Casana, and M.M.\ Ferreira,
Phys.\ Rev.\ D {\bf 92}, 025049 (2015);
Phys.\ Lett.\ B {\bf 760}, 302 (2016).

\bibitem {Jonas3}
J.B.\ Araujo, R.\ Casana, and M.M.\ Ferreira,
Phys.\ Rev.\ D {\bf 97}, 055032 (2018).

\bibitem {NMC}
G.\ Gazzola \etal,
J.\ Phys.\ G {\bf 39}, 035002 (2012);
L.C.T.\ Brito, H.G.\ Fargnoli, and A.P.\ Baêta Scarpelli,
Phys.\ Rev.\ D {\bf 87}, no. 12, 125023 (2013);
L.H.C.~Borges \etal, 
Phys.\ Lett.\ B {\bf 756}, 332 (2016);
Y.M.P.~Gomes and J.T.~Guaitolini Junior,
Phys.\ Rev.\ D {\bf 99}, 055006 (2019);
V.E.\ Mouchrek-Santos and M.M.\ Ferreira,
Phys.\ Rev.\ D {\bf 95}, no. 7, 071701 (2017);
J.\ Phys.\ Conf.\ Ser.\ {\bf 952}, 012019 (2018).

\bibitem{LVArbDim} 
V.A.\ Kosteleck\'y and Z.\ Li,
Phys.\ Rev.\ D {\bf 99}, 056016 (2019).

\bibitem {Yamanaka}
N.\ Yamanaka,
Int.\ J.\ Mod.\ Phys.\ E {\bf 26}, 1730002 (2017).

\bibitem {LeptonEDM}
B.L.\ Roberts and W.J.\ Marciano,
Adv.\ Ser.\ Direct.\ High Energy Phys.\ {\bf 20},~1 (2009).

\bibitem {EDMnature2018}
ACME Collaboration, V.\ Andreev \etal,
Nature {\bf 562}, 355 (2018).

\bibitem {added}
J.B.\ Araujo \etal, 
arXiv:1902.10329, 
submitted for publication in Phys.\ Rev.~D.

\end{thebibliography}
\end{document}